\documentclass[]{spie}  %>>> use for US letter paper
\usepackage{amsmath,amsfonts,amssymb}
\usepackage{graphicx}
\usepackage{multirow}
\usepackage[colorlinks=true, allcolors=blue]{hyperref}
\usepackage{cite}

\title{HAAT: Hybrid Attention Aggregation Transformer for Image Super-Resolution}

\author[a,b]{Song-Jiang Lai}
\author[a,b]{Tsun-Hin Cheung}
\author[a,b]{Ka-Chun Fung}
\author[a,b]{Kai-wen Xue}
\author[a,b]{Kin-Man Lam}
\affil[a]{Centre for Advances in Reliability and Safety. New Territories, Hong Kong}
\affil[b]{Department of Electrical and Electronic Engineering, The Hong Kong Polytechnic University. Kowloon, Hong Kong}

\authorinfo{Further author information: (Send correspondence to Songjiang-Lai.)\\Song-Jiang Lai.: E-mail: songjiang.Lai@connect.polyu.hk\\  Tsun-Hin Cheung.: E-mail: tsun-hin.cheung@connect.polyu.hk}

\begin{document}
\maketitle

\begin{abstract}
In the research area of image super-resolution, Swin-transformer-based models are favored for their global spatial modeling and shifting window attention mechanism. However, existing methods often limit self-attention to non-overlapping windows to cut costs and ignore the useful information that exists across channels. To address this issue, this paper introduces a novel model, the Hybrid Attention Aggregation Transformer (HAAT), designed to better leverage feature information. HAAT is constructed by integrating Swin-Dense-Residual-Connected Blocks (SDRCB) with Hybrid Grid Attention Blocks (HGAB). SDRCB expands the receptive field while maintaining a streamlined architecture, resulting in enhanced performance. HGAB incorporates channel attention, sparse attention, and window attention to improve nonlocal feature fusion and achieve more visually compelling results. Experimental evaluations demonstrate that HAAT surpasses state-of-the-art methods on benchmark datasets.    
\end{abstract}

% Include a list of keywords after the abstract 

\section{INTRODUCTION}
\label{sec:intro}  % \label{} allows reference to this section

% Single Image Super-Resolution (SISR) focuses on reconstructing a high-quality image from its low-resolution counterpart. Developing efficient super-resolution algorithms has become a key area of research in computer vision due to its broad applications. Recent studies have incorporated the self-attention mechanism into computer vision tasks\cite{liang2021swinir, liu2021swin}.

Single Image Super-Resolution (SISR) aims to reconstruct a high-quality image from a low-resolution one. The development of effective super-resolution algorithms has emerged as a pivotal research domain in computer vision owing to its extensive applications. Recent research has integrated the self-attention mechanism into computer vision challenges\cite{liang2021swinir, liu2021swin}.

CNN-based techniques for Single Image Super-Resolution (SISR) have markedly improved the restoration of image texture features. SRCNN\cite{yang2019deep} was the inaugural model to tackle super-resolution with convolutional neural networks. VDSR\cite{kim2016accurate} implemented residual learning to enhance learning and successfully address the gradient vanishing issue in deep networks. In SRGAN\cite{ledig2017photo}, Ledig et al. employed generative adversarial networks to refine super-resolution image generation, with the generator converting low-resolution images to high-resolution ones and improving quality via adversarial training. ESRGAN\cite{Hanson93c} included the Residual Dense Block (RRDB) as a fundamental network component, diminishing perceptual loss by utilizing characteristics prior to activation, hence producing images with more authentic textures. Moreover, researchers persist in suggesting novel structures to retrieve progressively realistic information in super-resolution images. CNN-based networks have demonstrated considerable performance efficacy. Nonetheless, the inductive bias inherent in CNNs constrains the ability of SISR models to capture long-range relationships. The constraints arise from the parameter-dependent scaling of the receptive field and the convolution operator's kernel size across many layers, potentially neglecting non-local spatial information in pictures.

To enhance the joint modeling of different hierarchical structures in pictures, researchers have taken advantage of the self-attention mechanism's benefits in multi-scale processing and long-range dependency modeling. Transformer-based SISR models have been developed to overcome the shortcomings of CNN-based networks by utilizing their capacity to simulate long-range dependencies and improving SISR performance. For instance, super-resolution results have been markedly enhanced by SwinIR\cite{liang2021swinir}, which makes use of the Swin Transformer\cite{liu2021swin}. Furthermore, state-of-the-art results have been produced with a hybrid attention transformer (HAT)\cite{chen2023activating} that combines an overlapping cross-attention module with window-based self-attention and channel attention.

% Although Transformer-based methods have been successfully applied to image restoration tasks, there are still some things that could be improved. Existing window-based Transformer networks restrict the self-attention computation to a dense area. This strategy obviously leads to a limited receptive field and does not fully utilize the feature information from the original image. In order to overcome the above problems, we propose a hybrid multiaxial aggregation network called HAAT in this paper. HAAT is built by combining Swin-Dense-Residual-Connected Blocks (SDRCB)\cite{hsu2024drct} with Hybrid Grid Attention Blocks (HGAB). HGAB inspired by GAB\cite{chu2024hmanet}, combines channel attention, sparse attention and window attention, which utilizes channel attention’s global information perception capability to compensate for self-attention’s short comings. The sparse self-attention is introduced to increase global feature interactions while balancing the computational complexity. Meanwhile, to further excite the potential performance of the model.

Despite the successful use of Transformer-based approaches to image restoration problems, there are areas for improvement. Current window-based Transformer networks confine self-attention calculations to a concentrated region. This method results in a constrained receptive field and fails to properly use the feature information from the original picture. This research proposes a hybrid multiaxial aggregation network, termed HAAT, to address these issues. HAAT is constructed by integrating Swin-Dense-Residual-Connected Blocks (SDRCB)\cite{hsu2024drct} with Hybrid Grid Attention Blocks (HGAB). HGAB, inspired by GAB\cite{chu2024hmanet}, integrates channel attention, sparse attention, and window attention, leveraging the global information perception capabilities of channel attention to address the deficiencies of self-attention. Sparse self-attention is used to enhance global feature interactions while maintaining computational efficiency further enhancing the potential performance of the model.

\section{HYBRID ATTENTION AGGREGATION TRANSFORMER}

\begin{figure} [ht]
   \begin{center}
   \begin{tabular}{c} %% tabular useful for creating an array of images 
   \includegraphics[height=7cm]{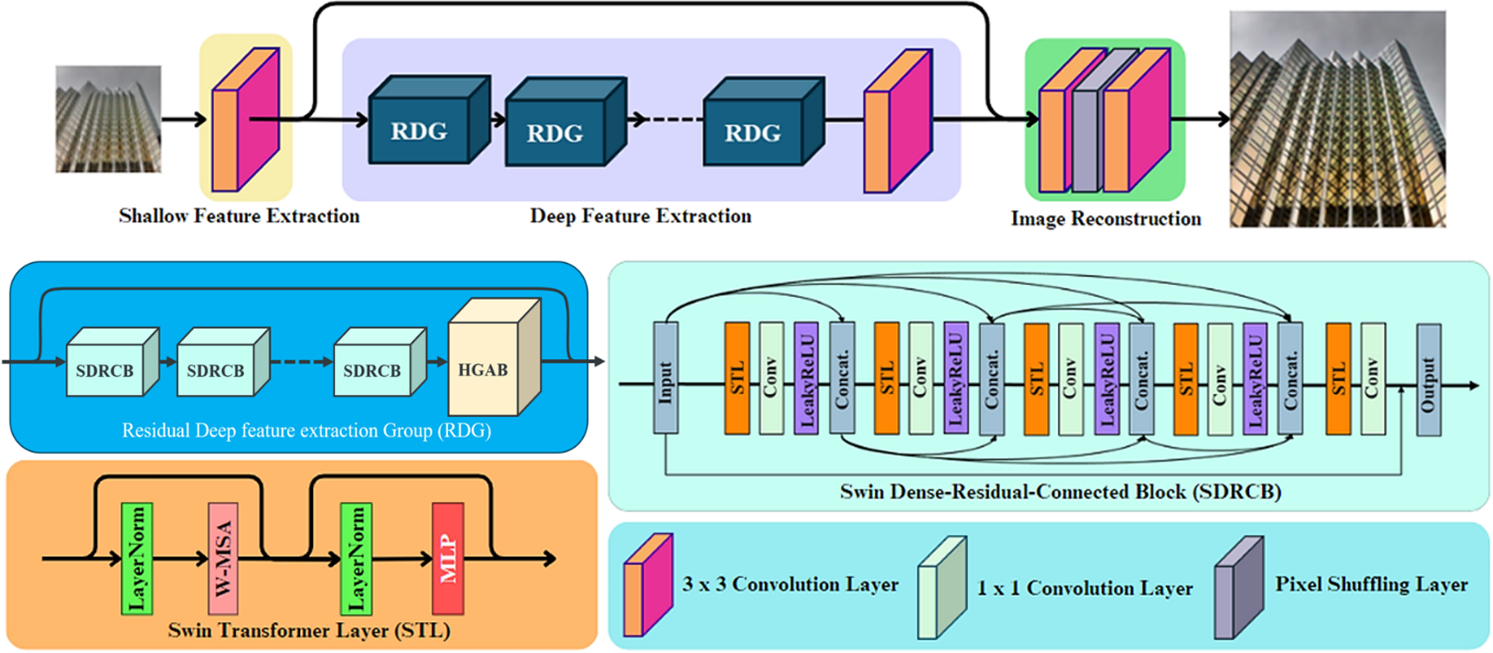}
   \end{tabular}
   \end{center}
   \caption[example] 
%>>>> use \label inside caption to get Fig. number with \ref{}
   { \label{fig:example} 
  SDRCB Framework.}
   \end{figure}

% The entire structure diagram of HAAT is illustrated in Fig.1. SDRCB incorporates Swin Transformer Layers and transition layers into each Residual Deep feature extraction Group (RDG). This approach enhances the receptive field with fewer parameters and a simplified model architecture, thereby resulting in improved performance. Besides, we introduce HGAB to model cross-area similarity for enhanced image reconstruction. The structure of HGAB is shown as Fig.2. The HGAB consists of a Mix Attention Layer (MAL) and an MLP layer. The proposed HGAB adopts the idea of sparse self-attention to increase global feature interactions while balancing the computational complexity. Therefore, our method allows joint modeling using similar features to generate better reconstructed images. 

% The overall structure of HAAT is shown in Fig. 1. SDRCB integrates Swin Transformer Layers and transition layers into each Residual Deep feature extraction Group (RDG), expanding the receptive field with fewer parameters and a simpler architecture, leading to improved performance. Additionally, we introduce HGAB to model cross-area similarity for better image reconstruction. The structure of HGAB, shown in Fig. 2, includes a Mix Attention Layer (MAL) and an MLP layer. HGAB uses sparse self-attention to enhance global feature interactions while managing computational complexity, enabling joint modeling of similar features for improved image reconstruction.

Figure 1 illustrates the comprehensive structure of HAAT. SDRCB incorporates Swin Transformer Layers and transition layers into each Residual Deep feature extraction Group (RDG), enhancing the receptive field while using fewer parameters and a more streamlined design, resulting in superior performance. Furthermore, we provide HGAB to describe cross-area similarity for enhanced picture reconstruction. The architecture of HGAB, seen in Figure 2, comprises a Mix Attention Layer (MAL) and a Multi-Layer Perceptron (MLP) layer. HGAB employs sparse self-attention to augment global feature interactions while controlling computational complexity, facilitating the joint modeling of analogous features for enhanced picture reconstruction. Furthermore, the employed channel attention mechanism can help the model extract more effective information between different channels.

% \begin{figure} [ht]
%    \begin{center}
%    \begin{tabular}{c} %% tabular useful for creating an array of images 
%    \includegraphics[height=6cm]{HGAB.png}
%    \end{tabular}
%    \end{center}
%    \caption[example] 
% %>>>> use \label inside caption to get Fig. number with \ref{}
%    { \label{fig:example} 
%   HGAB Structure.}
%    \end{figure}

\subsection{Swin-Dense-Residual-Connected Block}
\label{sec:title}

We use the shifting window self-attention mechanism of the Swin-Transformer Layer (STL)\cite{liang2021swinir, liu2021swin} to capture long-range dependencies via adaptive receptive fields. STL modifies the model's emphasis according to global content, enhancing feature extraction. This technique maintains global details as the network deepens, enlarging the receptive area without degradation. Integrating STL with dense-residual connections expands the receptive area and improves emphasis on critical information, hence increasing performance in SISR tasks that need thorough, context-sensitive processing. The SDRCB for input feature maps \( \boldsymbol{Z} \) inside RDG is delineated as follows.

\begin{equation}
\label{eq:fov}
\boldsymbol{Z_j} = H_{trans}(\boldsymbol{STL}([\boldsymbol{Z},...\boldsymbol{Z_{j-1}}]),j=1,2,3,4,5,
\end{equation}

\begin{equation}
\label{eq:fov}
SDRCB(\boldsymbol{Z}) = \alpha \cdot \boldsymbol{Z_5} + \boldsymbol{Z},
\end{equation}

where $[\cdot]$ denotes the concatenation of multi-level feature maps produced by the previous layers. $H_{trans}(\cdot)$ refers to the convolution layer with a LeakyReLU activation function for feature transition. The negative slope of LeakyReLU is set to 0.2. Conv1 is the $1\times1$ convolution layer, which is used to adaptively fuse a range of features with different levels\cite{tai2017memnet}. $\alpha$ represents residual scaling factor, which is set to 0.2 for stabilizing the training process\cite{Hanson93c}.

\subsection{Hybrid Grid Attention Block(HGAB)}

\begin{figure} [ht]
   \begin{center}
   \begin{tabular}{c} %% tabular useful for creating an array of images 
   \includegraphics[height=6cm]{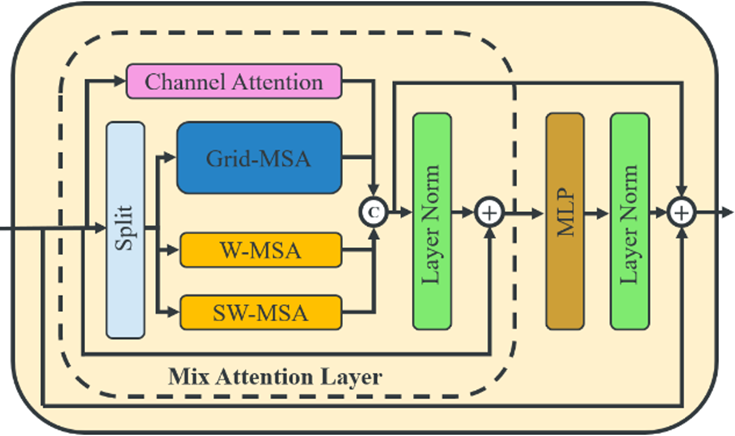}
   \end{tabular}
   \end{center}
   \caption[example] 
%>>>> use \label inside caption to get Fig. number with \ref{}
   { \label{fig:example} 
 The structure of HGAB.}
   \end{figure}

The structure of Hybrid Grid Attention Block (HGAB) is illustrated in Figure 2. This advanced hybrid attention block integrates channel, sparse, and self-attention to enhance feature modeling and representation learning beyond traditional hybrid attention approaches. Within the Hybrid Global Attention Block (HGAB), the Multi-Attention Layer (MAL) enables parallel processing of divided input features, boosting efficiency. This design improves global feature interaction and utilizes non-local spatial information, overcoming self-attention limitations while controlling computational complexity. Consequently, it significantly enhances performance in image super-resolution tasks through more sophisticated feature extraction and learning. The HGAB consists of a Mix Attention Layer (MAL) and an MLP layer. Regarding the MAL, we first split the input feature $F_{in}$ into two parts by channel: $F_G \in R^{(H \times W \times {C/2})}$ and $F_W \in R^{(H \times W \times {C/2})}$. Besides, the input is fed to another branch, where channel attention s applied. Subsequently, we split $F_W$ into two parts by channel again and input them into W-MSA and SW-MSA, respectively. Meanwhile, $F_G$ is input into Grid-MSA\cite{chu2024hmanet}.The computation process of MAL is as follows:

\begin{equation}
\label{eq:fov}
X_{W_1} = W - MSA(F_{W_1}),
\end{equation}

\begin{equation}
\label{eq:fov}
X_{W_2} = SW - MSA(F_{W_2}),
\end{equation}

\begin{equation}
\label{eq:fov}
X_G = Grid - MSA(F_G),
\end{equation}

\begin{equation}
\label{eq:fov}
X_C = CA(F_{in}),
\end{equation}

\begin{equation}
\label{eq:fov}
X_{MAL} = LN(Cat(X_{W_1},X_{W_2},X_{W_G})+X_C)+F_{in},
\end{equation}

where $X_{W_1}$, $X_{W_2}$, $X_G$ and $X_C$ are the output features of W-MSA, SW-MSA, Grid-MSA, and CA, respectively. Furthermore, it is worth noting that we adopt the post-norm method in HGAB to enhance the network training stability. For a given input feature $F_{in}$, the computation process of HGAB is as follows:

\begin{equation}
\label{eq:fov}
F_M = LN(MAL(F_{in}))+F_{in},
\end{equation}

\begin{equation}
\label{eq:fov}
F_M = LN(MAL(F_M))+F_M,
\end{equation}

\section{EXPERIMENTAL RESULTS}
\label{sec:sections}

\begin{table}[]
 \centering
 \caption{Quantitative comparison with SOTA methods}
\resizebox{0.7\columnwidth}{!}{%
\begin{tabular}{|c|c|c|c|c|c|c|} 
\hline
\multirow{2}{*}{\textbf{Method}} & \multirow{2}{*}{\textbf{Scale}} & \multirow{2}{*}{\textbf{Training Dataset}} & \multicolumn{2}{c|}{\textbf{Set5}} & \multicolumn{2}{c|}{\textbf{Set14}}  \\ 
\cline{4-7}
                                 &                                 &                                            & PSNR   & SSIM                      & PSNR   & SSIM                        \\ 
\hline
EDSR                             & ×2                              & DIV2K                                      & 38.11 & 0.9602                   & 33.92 & 0.9195                     \\
RCAN                             & ×2                              & DIV2K                                      & 38.27 & 0.9614                   & 34.12 & 0.9216                     \\
SAN                              & ×2                              & DIV2K                                      & 38.31 & 0.9620                   & 34.07 & 0.9213                     \\
IGNN                             & ×2                              & DIV2K                                      & 38.24 & 0.9613                   & 34.07 & 0.9217                     \\
HAN                              & ×2                              & DIV2K                                      & 38.27 & 0.9614                   & 34.16 & 0.9217                     \\
NLSN                             & ×2                              & DIV2K                                      & 38.34 & 0.9618                   & 34.08 & 0.9231                     \\
SwinIR                           & ×2                              & DIFK                                       & 38.42 & 0.9623                   & 34.46 & 0.9250                     \\
CAT-A                            & ×2                              & DIFK                                       & 38.51 & 0.9626                   & 34.78 & 0.9265                     \\
HAT                              & ×2                              & DIFK                                       & 38.63 & 0.9630                   & 34.86 & 0.9274                     \\
DAT                              & ×2                              & DIFK                                       & 38.58 & 0.9629                   & 34.81 & 0.9272                     \\
DRCT                             & ×2                              & DIFK                                       & 38.72 & 0.9646                   & 34.96 & 0.9287                     \\
\textbf{HAAT (Ours)}             & ×2                              & DIFK                                       & 38.74 & 0.9645                   & 34.97 & 0.9287                     \\ 
\hline
EDSR                             & ×3                              & DIV2K                                      & 34.65 & 0.9280                   & 30.52 & 0.8462                     \\
RCAN                             & ×3                              & DIV2K                                      & 34.74 & 0.9299                   & 30.65 & 0.8482                     \\
SAN                              & ×3                              & DIV2K                                      & 34.75 & 0.9300                   & 30.59 & 0.8476                     \\
IGNN                             & ×3                              & DIV2K                                      & 34.72 & 0.9298                   & 30.66 & 0.8484                     \\
HAN                              & ×3                              & DIV2K                                      & 34.75 & 0.9299                   & 30.67 & 0.8483                     \\
NLSN                             & ×3                              & DIV2K                                      & 34.85 & 0.9306                   & 30.70 & 0.8485                     \\
SwinIR                           & ×3                              & DIFK                                       & 34.97 & 0.9318                   & 30.93 & 0.8534                     \\
CAT-A                            & ×3                              & DIFK                                       & 35.06 & 0.9326                   & 31.04 & 0.8538                     \\
HAT                              & ×3                              & DIFK                                       & 35.07 & 0.9329                   & 31.08 & 0.8555                     \\
DAT                              & ×3                              & DIFK                                       & 35.16 & 0.9331                   & 31.11 & 0.8550                     \\
DRCT                             & ×3                              & DIFK                                       & 35.15 & 0.9333                   & 31.22 & 0.8569                     \\
\textbf{HAAT (Ours)}             & ×3                              & DIFK                                       & 35.17 & 0.9336                   & 31.23 & 0.8569                     \\ 
\hline
EDSR                             & ×4                              & DIV2K                                      & 32.46 & 0.8968                   & 28.80 & 0.7876                     \\
RCAN                             & ×4                              & DIV2K                                      & 32.63 & 0.9002                   & 28.87 & 0.7889                     \\
SAN                              & ×4                              & DIV2K                                      & 32.64 & 0.9003                   & 28.92 & 0.7888                     \\
IGNN                             & ×4                              & DIV2K                                      & 32.57 & 0.8998                   & 28.85 & 0.7891                     \\
HAN                              & ×4                              & DIV2K                                      & 32.64 & 0.9002                   & 28.90 & 0.7890                     \\
NLSN                             & ×4                              & DIV2K                                      & 32.59 & 0.9000                   & 28.87 & 0.7891                     \\
SwinIR                           & ×4                              & DIFK                                       & 32.92 & 0.9044                   & 29.09 & 0.7950                     \\
CAT-A                            & ×4                              & DIFK                                       & 33.08 & 0.9052                   & 29.18 & 0.7960                     \\
HAT                              & ×4                              & DIFK                                       & 33.04 & 0.9056                   & 29.23 & 0.7973                     \\
DAT                              & ×4                              & DIFK                                       & 33.08 & 0.9055                   & 29.23 & 0.7973                     \\
DRCT                             & ×4                              & DIFK                                       & 33.09 & 0.9061                   & 29.32 & 0.7982                     \\
\textbf{HAAT (Ours)}             & ×4                              & DIFK                                       & 33.12 & 0.9062                   & 29.32 & 0.7983                     \\
\hline
\end{tabular}
}
\end{table}

Our HAAT model is trained on DF2K, a substantial aggregated dataset that includes DIV2K\cite{agustsson2017ntire} and Flickr2K\cite{timofte2017ntire}. DIV2K provides 800 images for training, while Flickr2K contributes 2650 images. For the training input, we generate LR versions of these images by applying the bicubic down sampling method with scaling factors of 2, 3, and 4. To assess the effectiveness of our model, we conduct performance evaluations using well-known SISR benchmark datasets such as Set5\cite{bevilacqua2012low} and Set14\cite{zeyde2012single}.

In the DRCT architecture, the depth and width configuration mirrors that of HAT. Specifically, both models have 6 RDG and SDRCB units, with 180 channels for intermediate feature maps. For window-based multi-head self-attention (W-MSA), the number of attention heads is set to 6, and the window size is 16. In the HGAB block, the channel squeeze factor is 16, with 180 channels for intermediate features. The Grid MSA and (S)W-MSA use 3 and 2 attention heads, respectively. HR patches of 256 × 256 pixels were extracted from HR images, with random horizontal flips and rotation for data augmentation. As shown in Table 1, our method outperforms state-of-the-art techniques in both PSNR and SSIM.

For evaluation, we use all RGB channels and exclude the outermost (2 × scale) border pixels. PSNR and SSIM metrics are employed for assessment. Table 1 shows a quantitative comparison of our method with state-of-the-art approaches such as EDSR\cite{lim2017enhanced}, RCAN\cite{zhang2018image}, SAN\cite{dai2019second}, IGN\cite{zhou2020cross}, HAN\cite{niu2020single}, NLSN\cite{mei2021image}, SwinIR \cite{liang2021swinir}, CATA\cite{chen2022cross}, DAT\cite{chen2023dual} and CDRT\cite{hsu2024drct}. Our method consistently outperforms these methods across all benchmark datasets. Therefore, it can be clearly summarized HAAT achieves significantly better results than the other state-of-the-art models.

\section{CONCLUSION}
\label{sec:sections}

This work introduces a unique Hybrid Attention Aggregation Transformer (HAAT) for single-image super-resolution. HAAT enhances the DRCT architecture, emphasizing the stabilization of information flow and the expansion of receptive fields via dense connections in residual blocks, in conjunction with the shift-window attention mechanism to adaptively acquire global information. This enables the model to enhance its emphasis on global geographical information, optimizing its capabilities and circumventing information bottlenecks. Furthermore, motivated by the hierarchical structural resemblance in images, we provide HGAB to represent long-range relationships. The network improves multi-level structural similarity via the integration of channel attention, sparse attention, and window attention. The model was trained on the DF2K dataset and verified using the Set5 and Set14 datasets. Experimental findings indicate that our strategy surpasses SOTA techniques on benchmark datasets for single-image super-resolution tasks.

% \subsection{Subsection Attributes}

% The subsection heading is left justified and set in 11-point, bold font.  Capitalization rules are the same as those for book titles.  The first word of a subsection heading is capitalized.  The remaining words are also capitalized, except for minor words with fewer than four letters, such as articles (a, an, and the), short prepositions (of, at, by, for, in, etc.), and short conjunctions (and, or, as, but, etc.).  Subsection numbers consist of the section number, followed by a period, and the subsection number within that section.  

% \subsubsection{Sub-subsection attributes}
% The sub-subsection heading is left justified and its font is 10 point, bold.  Capitalize as for sentences.  The first word of a sub-subsection heading is capitalized.  The rest of the heading is not capitalized, except for acronyms and proper names.  

% References
\bibliography{report} % bibliography data in report.bib
\bibliographystyle{spiebib} % makes bibtex use spiebib.bst
\end{document}